\documentclass[12pt,dvips]{article}
\usepackage{amscd,verbatim}
\usepackage[all]{xy}
\usepackage{graphicx}

\usepackage{amsmath}
\usepackage[psamsfonts]{amssymb}
\usepackage{amsthm}
\usepackage{euscript}

\usepackage{latexsym}

\renewcommand{\(}{\begin{equation}}
\renewcommand{\)}{end{equation} \vspace{-.05in}\linebreak}

\newcounter{saveeqn}
\newcounter{savealpheqn}

\newcommand{\alpheqn}{\setcounter{saveeqn}{\value{equation}}%
  \stepcounter{saveeqn}\setcounter{equation}{0}%
  \renewcommand{\theequation}{\mbox{\arabic{section}.\arabic{saveeqn}
\alph{equation}}}
  \renewcommand{\)}{\end{equation}}}
\def\part#1{\frac{\partial}{\partial{#1}}}%
\def\group#1{\refstepcounter{equation}\setcounter{saveeqn}{\value{equati
on}}%
  \label{#1}\setcounter{equation}{0}%
\renewcommand{\theequation}{\mbox{\arabic{section}.\arabic{saveeqn}
\alph{equation}}}
  \renewcommand{\)}{\end{equation}}}
\newcommand{\reseteqn}{\setcounter{equation}{\value{saveeqn}}%
  \renewcommand{\theequation}{\arabic{section}.\arabic{equation}}%
  \renewcommand{\)}{\end{equation}}}

\newcommand{\aalpheqn}{\setcounter{saveeqn}{\value{equation}}%
  \stepcounter{saveeqn}\setcounter{equation}{0}%
  \renewcommand{\theequation}{\mbox{
        \Alph{subsection}.\arabic{saveeqn}\alph{equation}}}
   \renewcommand{\)}{\end{equation}}}
\newcommand{\areseteqn}{\setcounter{equation}{\value{saveeqn}}%
  \renewcommand{\theequation}{\Alph{subsection}.\arabic{equation}}%
  \renewcommand{\)}{\end{equation}}}

\renewcommand{\thefootnote}{\alph{footnote}}
\renewcommand{\(}{\begin{equation}}
\renewcommand{\)}{\end{equation}}
\newcommand{\ba}{\begin{eqnarray}}
\newcommand{\ea}{\end{eqnarray}}

\newcommand{\bp}{\mathop{\vtop{\ialign{##\crcr
   $\hfil\displaystyle{}\hfil$\crcr\noalign{\kern-13pt\nointerlineskip}
   \BIG{(}\hskip0pt\crcr\noalign{\kern3pt}}}}}
\newcommand{\cbp}{\mathop{\vtop{\ialign{##\crcr
   $\hfil\displaystyle{}\hfil$\crcr\noalign{\kern-13pt\nointerlineskip}
   \BIG{)}\hskip0pt\crcr\noalign{\kern3pt}}}}}
\newcommand{\pa}{\mathop{\vtop{\ialign{##\crcr

$\hfil\displaystyle{\oplus}\hfil$\crcr\noalign{\kern+1pt\nointerlineskip
}
   \hspace{.08in}$^{\alpha=0}$\hskip6pt\crcr\noalign{\kern3pt}}}}}

\newcommand{\beq}{\begin{equation}}
\newcommand{\eeq}{\end{equation}}

\numberwithin{equation}{section}

\def\hsp#1{\hspace{#1in}}

\catcode`\@=11
\def\vereq#1#2{\lower3pt\vbox{\baselineskip1.5pt \lineskip1.5pt
\ialign{$\m@th#1\hfill##\hfil$\crcr#2\crcr\sim\crcr}}}
\catcode`\@=12

\makeatletter
\newcommand\figcaption{\def\@captype{figure}\caption}
\newcommand\tabcaption{\def\@captype{table}\caption}
\makeatother
\renewcommand{\(}{\begin{equation}}
\renewcommand{\)}{\end{equation}}

\newcommand{\RR}{{\mathbb R}}
\newcommand{\ZZ}{{\mathbb Z}}

\theoremstyle{plain}

\theoremstyle{definition}

\begin{document}

\begin{titlepage}
\begin{flushright}

hep-th/0507106
\end{flushright}

\vspace{2em}
\def\thefootnote{\fnsymbol{footnote}}

\begin{center}
{\Large\bf Flux quantization and the M-theoretic characters}
\end{center}
\vspace{1em}

\begin{center}
{\large Hisham Sati} \footnote{E-mail: {\tt hisham.sati@adelaide.edu.au}\\
Research supported by the Australian Research Council. }
\end{center}

\begin{center}
\vspace{1em} {\em {
Department of Physics\\
and\\
Department of Pure Mathematics\\
       University of Adelaide\\
       Adelaide, SA 5005,\\
       Australia\\
\hsp{.3}\\
Department of Theoretical Physics\\
Research School of Physical Sciences and Engineering\\
The Australian National University\\
Canberra, ACT 0200\\
Australia}}\\

\end{center}

\vspace{0em}
\begin{abstract}
\noindent In a previous work \cite{S} we introduced characters and
classes built out of the M-theory four-form and the Pontrjagin
classes, which we used to express the Chern-Simons and the
one-loop terms in a way that makes the topological structures
behind them more transparent. In this paper we further investigate
such classes and the corresponding candidate generalized
cohomology theories. In particular, we study the flux quantization
conditions that arise in this context.

\end{abstract}

\vfill

\end{titlepage}
\setcounter{footnote}{0}
\renewcommand{\thefootnote}{\arabic{footnote}}

\pagebreak
\renewcommand{\thepage}{\arabic{page}}

\section{Introduction and discussion}

M-theory (see e.g. \cite{T}, \cite{D} for reviews) is the theory
in eleven dimensions which is believed to unify all the consistent
superstring theories and hoped to be the unified theory of quantum
gravity. A formulation of such theory seems to be unattainable at
present. However, one can hope to understand the theory from
various limits and using tools that continue to hold in the `bulk
of the moduli space' of the theory. Examples include studying the
BPS objects, such as the membrane and the fivebrane, and
investigating the structure of the topological terms in the
Lagrangian, namely the Chern-Simons term and the one-loop term
\cite{Loop}.

\vspace{3mm} Eleven-dimensional supergravity \cite{CJS} has three
fields in its supermultiplet, a metric $g$, a Rarita-Schwinger
field $\psi_{\mu}$ and a three-form $C_3$. The latter is
responsible for non-preturbative effects such as instantons, and
so in some sense encodes the topology of the theory (in the
Euclidean signature). In cases when cohomology is not an issue, we
have $G_4=dC_3$ as the `field strength'. In addition to $G_4$, one
can also have its (Hodge) dual $*G_4$, which is a rank seven
field.

\vspace{3mm} It was shown by Witten \cite{Wi1} that part of the
action is encoded by index theory of an $E_8$ bundle as well as a
of a Rarita-Schwinger bundle. The physical nature of the first of
the two bundles is not yet completely understood. Motivated by the
question of whether this theory is physical, i.e. having physical
degrees of freedom, \cite{ES} started an investigation of
supersymmetry, using an approximate model for $C_3$. Later,
\cite{DFM} gave a more accurate model \footnote{In fact they give
more than one.} which gave some insight into the nature of the
$E_8$ theory.

\vspace{3mm} M-theory is the strong coupling limit of ten
dimensional type IIA superstring theory \cite{various}. The
spectrum of the latter theory contains the Ramond-Ramond (RR)
fields as well as the Neveu-Schwarz (NSNS) fields. The first set
of fields live in K-theory \cite{MW}, and when the second set is
included, the fields are described by twisted K-theory.

\vspace{3mm} The partition function of a system is an interesting
quantum-mechanical function that gives us information about e.g.
amplitudes for certain transitions to occur, and so is one of the
most basic useful functions to calculate. The calculation of the
K-theoretic partition function for the RR fields in type IIA
superstring theory was started in \cite{MW} and completed in
\cite{DMW}, where is was also compared to the one obtained from
M-theory. The fields of the latter live in cohomology, and so the
partition function involves the cohomology Jacobian. Further, the
partition function on M-theory with boundaries was studied in
\cite{DFM} and \cite{FM}. Adding the NSNS fields in leads to
twisted K-theory. Some aspects of the resulting partition function
in this case were studied in \cite{relations}.

\vspace{3mm} A powerful concept that helps us connect the
(seemingly) different theories together is that of dualities. Any
formalism which describes the structures in one theory should be
compatible with that describing those in the dual theory. The
compatibility of the K-theoretic description of the fields of type
II supersting theories in ten dimensions with T-duality was
studied in \cite{MS} and with S-duality in \cite{flux} \cite{KS2}.

\vspace{3mm} One can ask whether the mathematical theories
describing the fields above are in fact just K-theory and $E_8$
gauge theory, or whether one can find more refined theories in the
case of the former and a more transparent structure in the case of
the latter that can for example explain the origin of the
mysterious $E_8$ gauge theory.

\vspace{3mm} Motivated by canceling the Diaconescu-Moore-Witten
anomaly \cite{DMW}, a proposal was given in \cite{KS1} in which
K-theory is replaced by another generalized cohomology, namely
elliptic cohomology. The point was that in this theory, the
anomaly given by the seventh integral Stiefel-Whitney class is
absent since this is the obstruction to the orientability with
respect to elliptic cohomology in the same way as the third
integral Stiefel-Whitney class is the obstruction to orientability
of bundles with respect to K-theory. A construction of part of the
the partition function, namely the mod 2 part, of type IIA was
given in \cite {KS1}. Another motivation for elliptic cohomology
was the fact that the (generically) twisted K-theories cannot be
compatible with S-duality in type IIB string theory \cite{KS2}. A
proposal for interpreting the elliptic curve via a unified view of
both type II theories in F-theory, as well as modularity, was
given in \cite{KS3}.

\vspace{3mm} In M-theory one can ask the same question. Is the
theory described just by usual cohomology or is there a deeper
structure hidden somewhere. While there is no obvious technical
reason for doing this, unlike the case of IIA above, there is a
compelling conceptual reason which is to understand M-theory
itself. In particular, a first step would be perhaps to understand
the various aspects of the $C_3$ field and the physical and
topological effects resulting from it, as well as the
corresponding $E_8$ gauge theory. Besides, since in type II it was
found that cohomology is not adequate to describe many subtle
situations and that K-theory was needed, it would not be
surprising that the same line of thought would work for M-theory
as well. What one gains out of this enhancement of structure is
that while usual cohomology is perturbative, K-theory is
inherently nonperturbative \cite{A}. So, likewise, in M-theory, we
might ask for a `more nonperturbative' mathematical theory than
usual cohomology.

\vspace{3mm} From the structure of the Chern-Simons and the
one-loop terms in the action of M-theory \footnote{We do not know
what the action of M-theory is, so what we mean is really the
action of eleven-dimensional supergravity continued to M-theory.}
certain multiplicative characters were proposed in \cite{S} which
resemble the Chern character, but which are instead built out of
the M-theory four-form $G_4$ and the Pontrjagin classes $p_i$. So
in some sense, we are looking at structures generalizing vectors
bundles whose classes are the string classes, the degree one piece
of which is just the usual string class $\lambda=p_1/2$. In
\cite{S} it was further proposed that a theory of characteristic
classes based on $p_i$ or $\lambda_i$, instead of the Chern
classes for vector bundles, seems to be suggested by the above
structure in M-theory.

\vspace{3mm} One aspect of the story in \cite{S} was parity. In
order to get the correct expressions for the Chern-Simons plus the
one-loop terms in the action, a parity condition was imposed such
that only the parity-odd part (i.e. the terms containing an odd
number of $G_4$'s) was retained. From the point of view of the
gravitational characters, i.e. the ones whose degree eight
components gives the one-loop polynomial, this is a restriction on
the degrees. If we were to generalize this to higher dimensions,
then effectively this means keeping degrees $8k$ (and in our
physical case, $k=1$). This seems to be similar to getting the
Pontrjagin classes in terms of the Chern classes with only the
even degrees (i.e. real degree $4n$) contributing. So our
situation seems to be a higher dimensional analog of this, where
degrees $8k$ are obtained and degrees $8k+4$ are killed. Of
course, from the mathematical point of view there is a priori
nothing special about stopping at dimension 12. So in principle,
the mathematical side of our formalism should work as well in
higher dimensions.

\vspace{3mm} In this paper we follow the program above to see
whether one can gain insight about the structures related to the
non-gravitational fields in M-theory. At the classical
supergravity level, these are just differential forms. In the
quantum theory, one would then have thought that they would simply
be integral cohomology classes. It turns out that this is naive
and the actual story is much more subtle due to the presence of
the fermions. The contribution to the path integral from the
latter fields should be taken into account, and this leads to a
nontrivial shift in the quantization condition for $G_4$
\cite{Wi1}. One natural question is then whether one can get the
quantization conditions on these fields in a somewhat natural way
in the context of the characters introduced in \cite{S}. Together
with providing some further explanations relating to the
characters, this is the main theme of the current paper.

\section{Flux quantization}
We are guided by the quantization law on the four-form that was
derived by Witten \cite{Wi1}. This is of the form $G_4=a
-\frac{1}{2} \lambda $. Now there is something very interesting in
the structure of this formula, namely the factor of one-half in
the gravitational part of the expression. We find this interesting
for several reasons and we believe it connects nicely with other
aspects of M-theory and string theory. For example, the
topological part of the action of M-theory, composed of the
Chern-Simons and the one-loop, was shown by Witten to be written
in terms of the index of an $E_8$ bundle and that of a
Rarita-Schwinger bundle. The point is that the indices appear with
a relative factor of half again. In another instance, namely in
type II string theory, the Ramond-Ramond fields have a
quantization condition coming from K-theory, and there the
gravitational part also comes with a factor of half, except that
this time it is a square root (of the A-roof genus). The factor
appears as a $\frac{1}{2}$ in the additive structures and as a
power of half in the multiplicative structures. This
correspondence between multiplicative and additive is basically
what we are advocating in terms of exponentiating.

\vspace{3mm} Motivated by all the above, one could try to give the
simplest possible quantization condition that would put the
four-form within the framework of generalized cohomology theories
in the spirit of \cite{S} , namely a simple exponential of the
form \( G=e^{a(E_8)} \sqrt{ e^{\lambda(TX)}}. \label{prop}\) This
is attractive as it correctly produces the quantization condition
for $G_4$ when the degree four component is picked, and also seems
to be very analogous to the K-theoretic quantization formula for
the RR fields, i.e., \( F(x)=ch(x)
\sqrt{\widehat{A}(X)}. \) However there are several points to be
made and ambiguities to be explained here. First, if one is only
concerned with the four-form, then there is an ambiguity in
whether one chooses $e^{\lambda/2}$ in place of the square root
above, which would give the same answer for the four-form, but of
course will differ by increasing factors of one-half as one goes
up in degrees. The second is that the choice of sign in the
quantization condition, i.e. whether we pick the $\pm$ in $G_4=a
\pm \frac{1}{2}\lambda$ is reflected in the exponential in $e^{\pm
\lambda}$. Of course, $G_4$ by itself does not seem to be able to
pin down what multiplicative structure one can have. Thus we need
to go up in degrees.

\vspace{3mm} So then one can ask: what about the dual field
$*G_4$? Here we are first faced with the problem of trying to
identify the precise field that we would like to quantize.
Obviously, $*_{11}G_4$ has rank seven. This is not very attractive
from a mathematical point of view, as we would like to work with
an even degree field. The physics, however, can help us here. In
particular, the equation of motion for $G_4$, \(
d*G_4=-\frac{1}{2}G_4 \wedge G_4 + I_8 \) seems to tell us that
the `field' related to $*G_4$ that we need to look at is instead
the LHS of the equation of motion. Indeed such an expression was
studied by Diaconescu-Freed-Moore \cite{DFM} both in de Rham and
in integral cohomology, where there it was denoted $\Theta_X$.
Obviously, such an expression is very rich as it encodes powers of
$G_4$ as well as the one-loop gravitational polynomial
$I_8=\frac{p_2 -(p_1/2)^2}{48}$ in the Pontrjagin classes of the
manifold. So let us work with this degree eight expression
$\Theta$.

\vspace{3mm} Note that since
$[\frac{1}{2}G_4^2]=\frac{1}{2}(a-\lambda_1/2)^2$ then the class
$[\Theta_X(a)]\equiv [\frac{1}{2}G_4^2 - I_8]$ is equal to \(
\frac{1}{2}a^2 -\frac{1}{2}a\lambda_1 + \frac{1}{8}\lambda_1^2
-I_8. \label{thet} \) Since we will also be dealing with the
higher classes $\lambda_i$ which we defined in \cite{S}, we will
use $\lambda_1$ to denote the usual string (four) class in what follows.
One can group the four factors in the RHS of (\ref{thet}) in pairs
using that $30{\widehat A}_8 =\frac{1}{8}\lambda_1^2-I_8$ and
factorizing the first pair, to get \cite{DFM} \(
[\Theta_X(a)]=\frac{1}{2}a(a-\lambda_1) + 30 {\widehat A}_8. \)

\vspace{3mm} We next seek an expression for $\Theta$ in terms of
the new characters. Since the expression for $\Theta$ involves
(the square of) $G_4$, then we can use the discussion on $G_4$,
together with the expression for the purely gravitational part in
terms of the total string characters \cite{S}, to find the
expression \( \Theta_X=\left[ \frac{1}{2}\left( e^a
\sqrt{e^{\lambda_1}}\right)^2\right]_{(8)}
-\left[1+\frac{1}{24}(e^{\lambda}-1)\right]_{(8)}. \) However, we
would like to have an expression that does not use that for $G_4$,
but directly gives the desired formula independently of $G_4$.
Note also that there seems to be some asymmetry between the two
summands in the above expressions, namely that the first only
involves degree four classes (and in particular $\lambda_1$),
while the second involves the total string class, i.e. all degrees
$4k$ (in our case $k\leq 2$). If we were to insist to use the
above expression by putting $\lambda$ in place of $\lambda_1$ then
we would either obviously get extra unwanted factors or be forced
to impose conditions such as $\lambda_2=0$, which seem to be too
restrictive and somewhat unnatural.

\vspace{3mm} So then let us try to use an expression that contains
the total string class. From the proposed formula (\ref{prop}) for
the quantization of $G_4$ we are led to the analogous form
\begin{eqnarray}
\left[e^a e^{-\lambda}\right]_{(8)}&=&\frac{1}{2}a^2
-\frac{1}{2}a\lambda_1 +\frac{1}{4}(\lambda_1^2-2\lambda_2)
\nonumber\\
&=&\Theta - 30 {\widehat A}_8 -12 I_8
\nonumber\\
&=& \Theta -\frac{1}{8}\lambda_1^2 -11I_8 . \label{seek}
\end{eqnarray}
While this expression is suggestive, it is still not the final
form that we seek, because of the extra terms on its right hand
side.

\vspace{3mm} One can now ask whether it is reasonable to try to
seek a unified expression for both $G_4$ and $\Theta$ and  package
them together in a `total M-theory field'. This would be analogous
to writing the total Ramond-Ramond field compactly as $F=\sum_n
F_n$ where $n$ runs over all even (odd) numbers in type IIA (IIB)
with $n\leq 10$. The advantage of this in the case of type II was
obviously to say that it is the total RR field which is
collectively viewed as K-theoretic and that it does not make sense
to talk about the individual components in K-theory. For M-theory,
if an analogous construction goes through, then this would
similarly mean that one would talk about the `total M-theoretic
field' that lives in some generalized cohomology that we are
trying to identify.

 \vspace{3mm} Several aspects of the story
have to be taken into account in searching for a uniform formula
for the `total M-theory class' $G$. One important such concept 
is that of parity which also showed up earlier in
\cite{S} in picking the odd-parity part of the expression for the
topological part of the M-theory action, i.e. the Chern-Simons and
the one-loop terms. To connect with the above, we also require the
use of parity in the context of deriving the quantization
conditions for the four-form and its dual. We saw above that
certain terms need to be canceled (cf. (\ref{seek})) in order to
get a unified expression. Let us explore whether this is possible
with the aid of parity conditions. Of course, a natural way of
separating an exponential into parity components is to look at
hyberbolic sines and cosines. After all, this is how the A-genus
is constructed using the ${\rm sinh}$ function -- we look at some
aspects of this further in section \ref{gravity}. First note that
the term $\sqrt{e^{\pm \lambda}}$ does not factor nicely into
hyperbolic signs and cosines, and so this seems to suggest looking
at the other form $e^{\pm \lambda/2}$ which, as mentioned earlier,
is equivalent to the first form for the degree four part, but
differs by successive factors of halves in higher degrees. This is
basically the distinction between viewing the `half-ing' operation
as dividing the class by two then taking formal expansion vs.
taking the formal expansion in the undivided class and then taking
the formal square root of the resulting expression.

\vspace{3mm} Let us start with the expressions for the degree four
and eight components of $\sinh{(\lambda/2)}$. This gives
$\frac{1}{2}\lambda_1$ and $\frac{1}{2}\lambda_2$, respectively.
Similarly, for the components of $\cosh{(\lambda/2)}$, we
respectively have $0$ and $\frac{1}{8}\lambda_1^2$. Since we are
interested in multiplying those classes with the exponentiated
$E_8$ four-class $a$, we also need the expressions involving the
product with the hyperbolic sine

\begin{eqnarray}
e^a\sinh{(\lambda/2)}|_{(4)}&=&a + \frac{1}{2}\lambda_1
\nonumber\\
e^a\sinh{(\lambda/2)}|_{(8)}&=&\frac{1}{2}a^2 +
\frac{1}{2}a\lambda_1 +\frac{1}{2}\lambda_2,
\end{eqnarray}
and those with the hyperbolic cosine
\begin{eqnarray}
e^a\cosh{(\lambda/2)}|_{(4)}&=&a
\nonumber\\
e^a\cosh{(\lambda/2)}|_{(8)}&=&\frac{1}{2}a^2 +
\frac{1}{8}\lambda_1^2.
\end{eqnarray}

\vspace{3mm} But now note that the quantization conditions -- at
least that for $G_4$-- appear with some minus signs. How can this
be taken care of? This can be accommodated by the very simple
parity properties $\sinh(-x)=-\sinh(x)$ and $\cosh(-x)=\cosh(x)$.
This implies that we need to use the hyperbolic sine to get the
quantization of $G_4$, accounting for the minus sign. This is
simply obtained by \( e^a \sinh(\pm \lambda/2)|_{(4)}.
\label{approx} \)

\vspace{3mm} Then in order to check whether this makes sense for
the `total field strength', one has to look at the expression for
the eight form related to the dual field. If we use the same
expression, we have for the degree eight component \( e^a\sinh(\pm
\lambda/2)|_{(8)}=\frac{1}{2}a^2 \pm \frac{1}{2}a\lambda_1 \pm
\frac{1}{2}\lambda_2. \) At this stage let us see how far we are
from the desired expression. For this, we look at the difference
\( \Theta - e^a \sinh(-\lambda/2)=\frac{1}{8}\lambda_1^2 - I_8 +
\frac{1}{2}\lambda_2, \) so that regrouping gives \( \Theta -e^a
\sinh(-\lambda/2)+ I_8=\frac{1}{8}\lambda_1^2 +
\frac{1}{2}\lambda_2, \) which is just $e^{\lambda/2}|_{(8)}$.
\footnote{If we were to use a $\cosh$ term instead of an
exponential then we would not have been able to account for the
$\lambda_2$ term.} Therefore, we get the expression for the dual
class in terms of exponentiated degree eight classes \( \Theta
=e^a \sinh(-\lambda/2)|_{(8)}+ e^{\lambda/2}|_{(8)} - I_8. \) And
of course we already know the expression of $I_8$ in terms of the
string characters.

\vspace{3mm} Naturally, one can ask whether the expression just
derived correctly serves as the one for the `total field
strength'. Here, in going back to check whether the above
modification still respects the quantization condition for $G_4$,
we encounter a problem because the expression\( e^a
\sinh(-\lambda/2)|_{(4)}+ e^{\lambda/2}|_{(4)}=(a
-\frac{1}{2}\lambda_1) + \frac{1}{2}\lambda_1 \) gives only $a$
and cancels the gravitational term.

\vspace{3mm} This is not the end of the story, however. The above
suggests that, after all, perhaps we should confine ourselves to
four-classes only in dealing with the fields modulo the one-loop
term. Indeed, if we only use $a$ and $\lambda_1$, then we can
write the unified expression \(
G(a,\lambda_1)=e^a\sinh(-\lambda_1/2)+\cosh(-\lambda_1/2).
\label{main} \) Note that this is not very `far' from $e^a
e^{-\lambda_1/2}$, and in the absence of $a$ it is just
$e^{-\lambda_1/2}$. To check (\ref{main}), we evaluate the
four-form component to get \( G(a,\lambda_1)|_{(4)}=a
-\frac{1}{2}\lambda_1 \) which correctly reproduces the
quantization condition for $G_4$. Similarly, the eight-form
component gives \( G(a,\lambda_1)|_{(8)}=\frac{1}{2}a^2
-\frac{1}{2}a\lambda_1 +\frac{1}{8}\lambda_1^2, \) which correctly
reproduces the quantization condition for $\Theta+I_8$. Of course
there remains to incorporate the one-loop term into the story.

\section{More on the gravitational terms}
\label{gravity}
 First, we recall that the expression for the
components of the $A$-genus is given by the expansion of the
series $\prod \frac{z/2}{{\sinh(z/2)}}$ giving the usual
polynomials in the pontrjagin classes $p_i$. If we write those in
terms of the string classes $\lambda_i=p_i/2$ as defined in
\cite{S}, we get the following expression for $\widehat A$, \(
{\widehat A}=1-\frac{1}{12}\lambda_1 + \frac{1}{2^5.3^2.5}
\left(7\lambda_1^2 -2\lambda_2 \right)+
\frac{1}{2^7.3^3.5.7}\left( -31 \lambda_1^3 +22\lambda_1\lambda_2
- 4\lambda_3 \right) + \cdots . \)

\vspace{3mm} We will then look at the components, in each relevant
dimension, of the gravitational class introduced in \cite{S}. In
dimension four we have \( \left[1+\frac{1}{24}(e^{-\lambda}-1 )
\right]_{(4)}=-\frac{1}{24}\lambda. \) It is curious that this is
equal to $\left[\sqrt{\widehat{A}}\right]_{(4)}$. As for the
eight-form, it is designed to give the one-loop term
\begin{eqnarray}
\left[1+\frac{1}{24}(e^{-\lambda}-1 ) \right]_{(8)}&=&\frac{1}{24}
\left(-\lambda_2 +\frac{1}{2}(-\lambda_1)^2\right)
\nonumber\\
&=&\frac{-\lambda_2 + \frac{1}{2}(\lambda_1)^2}{24}
\nonumber\\
&=&-I_8.
\end{eqnarray}
For completeness, the degree twelve component is \(
\left[1+\frac{1}{24}(e^{-\lambda}-1 ) \right]_{(12)}=
\frac{1}{12}\left[\frac{1}{6}\lambda_1^3
-\frac{1}{2}\lambda_1\lambda_2 + \lambda_3 \right]. \)


\vspace{3mm} Let us write the new character in a slightly more
suggestive way to connect to the discussion on the total field
strength. Keeping both signs for generality, we write this as
follows \footnote{Recall that there are other variations on the
A-genus containing further hyperbolic functions, such as the
$G$-polynomial associated with the Rarita-Schwinger fields, which
has the expression \cite{AW} \( \prod _{n=1}^{D/2}
\frac{z_n/2}{\sinh{z_n/2}}\left( 2 \sum_{m=1}^{D/2} \cosh{z_m} -1
\right). \nonumber \)}
 \( 1+\frac{1}{24}(e^{\pm \lambda}-1
)=1+\frac{1}{12}e^{\pm\lambda/2}\sinh{(\pm\lambda/2)}.
\label{grav}\) The expression for the total field strength that
includes the one-loop term is then given by the sum of
(\ref{main}) and (\ref{grav}).

\vspace{3mm} As mentioned in the introduction, one interesting
aspect of the interpretation of the Ramond-Ramond fields in terms
of K-theory is the appearance of the $\sqrt{\widehat A}$ term.
Without it, of course we simply have the Chern character, which is
a map from K-theory to $\ZZ_2$-graded cohomology according to even
and odd degrees. There is a similar story in twisted K-theory
\cite{relations}, so that one has the map \( \sqrt{\widehat A}
\wedge ch_H: K(X,H) \mapsto H^{\rm even}(X,H), \) which is
actually an isomorphism over the rationals. The question is
whether there is an intrinsic mathematical interpretation of such
a factor. Freed and Hopkins \cite{FH} gave the following
interpretation. On compact ${\rm spin}^c$ manifolds, both $K^*(X)
\otimes \RR$ and $H^*(X;\RR)$ carry an addition, multiplication,
and a bilinear form. Physically, what seems to be relevant is the
addition-- for superposing states-- and the bilinear form-- for
electric or magnetic coupling-- but not the multiplication. The
modification of the Chern character, which preserves the addition
and multiplication but not the bilinear form, by $\sqrt{\widehat
A}$ preserves the desired properties, namely the addition and the
bilinear forms but not the multiplication.

\vspace{3mm} Another interpretation is given in the more general
context of generalized cohomology \cite{F}, where $\sqrt{\widehat
A}$ is a special choice of a normalizing differential form. In the
context of Riemannian fiber bundles, this appears in getting the
curvature on the base from the curvature on the total space of the
bundle, upon integration over the fiber. If $X \mapsto T$ is a
$\Gamma$-oriented fiber bundle (for generalized cohomology
$\Gamma$), then there is a closed differential form ${\widehat
A}_{\Gamma}(X/T)$ on $X$ so that if $\lambda \in
\Gamma^{\bullet}(X)$ has curvature $\omega$, then the curvature of
$\int_{X/T} \lambda$ is \( \int_{X/T} {\widehat
A}_{\Gamma}(X/T)\wedge \omega.  \) Then taking the square root is
a convenient choice as it makes bilinear pairings in $\Gamma$
compatible with integration of curvatures.

\vspace{3mm} In our case, we would like to think of the
gravitational terms (\ref{grav}) (after taking square roots as
appropriate) as normalizing forms in the proposed M-theoretic
generalized cohomology theory $\mathcal{M}$ whose character is
${\mathbb M}=e^{G_4}$ \cite{S}. If we call the term in
(\ref{grav}) $\sqrt {{\widehat A}_{\mathcal M}}$ then we have the
map \( \sqrt{{\widehat A}_{\mathcal M}} \wedge {\mathbb M} :
\mathcal{M} \longrightarrow H^{4k}. \) So how far is $\sqrt
{{\widehat A}_{\mathcal M}}$ from $\sqrt {\widehat A}$~ ? With the
above definition, we have agreement in degree four for the wedge
of either with ${\mathbb M}$ , as both give $G_4 +
\frac{1}{48}p_1$, but in dimension eight, we have \( \sqrt
{{\widehat A}_{\mathcal M}}\wedge {\mathbb
M}|_{(8)}=\frac{1}{2}G_4 \wedge G_4 - I_8 + G_4 \wedge
\frac{1}{24}\lambda_1. \) So in some sense, the formula involving
the new genus is in some sense a one-loop corrected formula for
the one containing the usual genus, and we may write schematically

\(  \sqrt {{\widehat A}_{\mathcal M}}\wedge{\mathbb M} = \sqrt
{\widehat A} \wedge {\mathbb M}~ + ~{\rm 1-loop}.\)

\vspace{3mm} Finally, for curiosity, let us see what happens if
the parity is not imposed, and see whether that leads to anything
as interesting. To keep the discussion as symmetric as possible,
let us use the expression (\ref{approx}) which uses the total
string class $\lambda$ instead of (\ref{main}) that uses the
degree four class only. Since we are not invoking parity in this
paragraph then for our purpose the difference is not very drastic.
If we add this time (\ref{grav}) to the term we found for the
total field strength, then we have the combination (choosing the
negative sign for $\lambda/2$) \( \left(e^a
+\frac{1}{12}e^{-\lambda/2} \right) \sinh(-\lambda/2). \label{tot}
\) We have mentioned several times above that we impose a parity
condition such that this `genus' does not contribute to the degree
four part of the field strength. However, let us consider what
happens had we not done that. The degree four component of
(\ref{tot}) would then give \( \left(
a+\frac{1}{12}\frac{-\lambda_1}{2}\right) \cdot 1 +
\frac{1}{12}\cdot (-\frac{1}{2}\lambda_1)=a
-\frac{1}{12}\lambda_1, \)  which is just $a+ {\widehat A}_1$. It
is interesting first that this is new quantization condition has
$p_1/24$ shift in it, as the $24$ is particularly curious because
this might be related to Topological Modular Forms, since the
obstruction to orientability with respect to TMF is in fourth
cohomology modulo $24\ZZ$ (cf. \cite{KS3}). Furthermore, it is
interesting that for the degree four components of the one-loop
term and of the total combination lead to the terms
$\sqrt{\widehat A}_1$ and ${\widehat A}_1$ respectively.


%
\end{document}